\documentclass[aps,prd,12pt]{revtex4}

\begin{document}
\def\eth{\epsilon}
\def\emx{\varepsilon}
\def\erf{\mbox{erf}}
\def\ER{E_{\rm R}}
\def\SD{{\rm SD}}
\def\SI{{\rm SI}}
\renewcommand{\baselinestretch}{1.0}
\title{Nuclear target effect on dark matter detection rate}     
\author{V.A.~Bednyakov}
\affiliation{Dzhelepov Laboratory of Nuclear Problems,
         Joint Institute for Nuclear Research, \\
         141980 Dubna, Russia; E-mail: Vadim.Bednyakov@jinr.ru}
\author{F. \v Simkovic}
\affiliation{Department of Nuclear Physics, Comenius University,
Mlynsk\'a dolina F1, SK--842 15 Bratislava, Slovakia}

\begin{abstract} 
	Expected event rates for 
	a number of dark matter nuclear targets were calculated
	in the effective low-energy minimal supersymmetric standard model, 
	provided the lightest neutralino is the dark matter 
	Weakly Interacting Massive Particle (WIMP).
	These calculations allow direct comparison of sensitivities 
	of different dark matter detectors to intermediate mass  
	WIMPs expected from the measurements of the DArk MAtter
	(DAMA) experiment.
\end{abstract} 

\maketitle 

\section{Introduction}
    Weakly Interacting Massive Particles (WIMPs) are among the
    most popular candidates for the relic cold dark matter (DM).
    There is some revival of interest to the 
    WIMP-nucleus spin-dependent interaction from both theoretical 
    (see e.g. 
\cite{Engel:1991wq,Bottino:2003cz,%
Bednyakov:1994te,Bednyakov:2000he,Bednyakov:2002mb,Bednyakov:2003wf,%
Bednyakov:2004qu,Bednyakov:2004xq,Bednyakov:2005qp})
     and experimental (see e.g. 
\cite{Girard:2005pt,Girard:2005dq,%
Giuliani:2004uk,Giuliani:2005bd,%
Savage:2004fn,Benoit:2004tt,Tanimori:2003xs,Ovchinnikov:2003AA,%
Moulin:2005sx,Mayet:2002ke,Klapdor-Kleingrothaus:2005rn})
      points of view.  
     There are some proposals aimed at direct DM detection with 
     relatively low-mass isotope targets 
\cite{Girard:2005pt,Girard:2005dq,Tanimori:2003xs,Ovchinnikov:2003AA,%
Moulin:2005sx,Mayet:2002ke}
      as well as some first attempts to design and construct 
      a DM detector which is sensitive to the nuclear recoil direction
\cite{Alner:2004cw,Snowden-Ifft:1999hz,Gaitskell:1996cv,Sekiya:2004ma,%
Morgan:2004ys,Vergados:2000cp,Vergados:2002bb}.
      Kinematically, low-mass targets 
      make preference for the low-mass WIMPs 
      (due to $M_{\rm Target}\approx M_{\rm WIMP}$) and 
      are more sensitive to the spin-dependent 
      WIMP-nucleus interaction as well 
\cite{Jungman:1996df,Engel:1991wq,Divari:2000dc,%
Bednyakov:1994te,Bednyakov:2000he,Bednyakov:2004xq,Bednyakov:1997ax}.
        On the other hand, WIMPs with masses about 100 GeV/$c^2$ 
	follow from the results of the DArk MAtter (DAMA) experiment.
	This collaboration claimed 
        observation of the first evidence for the dark matter signal
        due to registration of the annual modulation effect
\cite{Bernabei:2000qi,Bernabei:2003za,Bernabei:2003wy}.
	Aimed for more than one decade at 
	the DM particle direct detection, the DAMA experiment 
	with 100-kg highly radio-pure NaI(Tl) scintillator detectors  
	successfully operated till July 2002 
	at the Gran Sasso National Laboratory of the I.N.F.N.
	On the basis of the results obtained for over 7 annual cycles 
	(107731~kg$\cdot$day total exposure)
	the effectiveness of the WIMP model-independent 
	annual modulation signature was demonstrated 
	and the WIMP presence in the galactic halo is strongly supported 
	at 6.3 $\sigma$ C.L. 
\cite{Bernabei:2003za}.
	The main result of the DAMA observation 
	of the annual modulation signature 
	is the low-mass region of the WIMPs
	($40 < M^{}_{\rm WIMP} < 150$~GeV/$c^2$), provided these WIMPs are
	cold dark matter particles.
	It is obvious that such a serious claim should be verified 
	by  other independent measurements.  
	To reliably confirm or reject the DAMA result 
	an experiment should have 
	the same or better sensitivity to the annual modulation signal.
	At the same time one should know  an expected rate in his own
	detector, provided the DAMA result is correct. 
	Motivated by the DAMA evidence  
	predictions for the direct DM detection rate in a Ge-73 detector
	within the framework of the so-called 
	effective low-energy minimal supersymmetric standard
	model (effMSSM) are given in 
\cite{Bednyakov:2004be}. 
        Here the analysis of
\cite{Bednyakov:2004be} is extended and 
	the expected WIMP detection rates 
	are recalculated in the effMSSM
	for other DM-interesting isotope targets on the basis of 
	the data base from 
\cite{Bednyakov:2004be}. 
	 Similar expectations are obtained as well
	 by means of direct recalculations of the DAMA constraints 
\cite{Bernabei:2003za} 
          into detection rates for other targets.
	  This allows one to estimate the prospects 
	  to confirm or to reject the DAMA result by other 
	  DM search experiments. 

\section{Event rates and effective low-energy MSSM}  
        A dark matter event is an elastic scattering of a relic WIMP $\chi$ 
        from a target nucleus $A$ producing a nuclear 
        recoil $E_{\rm R}$ which can be detected by a suitable detector.
        The differential event rate in respect to the recoil 
        energy is the subject of experimental measurements.
        The rate depends on the distribution of the WIMPs 
	in the solar vicinity $f(v)$ and
        the cross section of WIMP-nucleus elastic scattering
\cite{Bernabei:2003za,Jungman:1996df,Lewin:1996rx,Smith:1990kw,%
Bottino:2000jx,%
Bednyakov:1999yr,Bednyakov:1996yt,Bednyakov:1997ax,Bednyakov:1997jr,%
Bednyakov:1994qa}.
        The differential event rate per unit mass of 
        the target material can be given as a sum of 
	spin-dependent (SD) and spin-independent (SI) contributions, 
	parameterized via 
	spin-independent ($\sigma^{}_\SI$) and spin-dependent ($\sigma^{}_\SD$) 
	WIMP-nucleon interaction cross sections
\cite{Bernabei:2003za,Bednyakov:2004be}: 
\begin{equation}
\label{Definitions.diff.rate}
\frac{dR(\ER)}{d\ER} 
	= N_T \frac{\rho_\chi}{m_\chi}
        \displaystyle
        \int^{v_{\rm max}}_{v_{\rm min}} dv f(v) v
        {\frac{d\sigma}{dq^2}} (v, q^2) 
	= 
          \kappa^{}_\SI(\ER,m_\chi)\,\sigma_\SI
         +\kappa^{}_\SD(\ER,m_\chi)\,\sigma_\SD.
\end{equation}
        The nuclear recoil energy
        $E_{\rm R} = q^2 /(2 M_A )$ is typically about $10^{-6} m_{\chi}$. 
        The number of nuclei per unit of target mass
        is $N_T$ and $M_A$ is the target nucleus mass.
	The effective spin WIMP-nucleon cross section
        $\sigma^{}_{\rm SD}$
        and the coupling mixing angle $\theta$ were introduced
\cite{Bernabei:2003za,Bernabei:2001ve}
        in such a way that SD WIMP-proton and SD
        WIMP-neutron interaction cross sections have the form
$\sigma^p_{\SD}=\sigma^{}_{\SD} \cdot \cos^2 \theta$, and 
$\sigma^n_{\SD}=\sigma^{}_{\SD} \cdot \sin^2 \theta$. 
        Further notations are:
\begin{eqnarray}
\nonumber
\kappa^{}_\SI(\ER,m_\chi)
&=&       N_T \frac{\rho_\chi M_A}{2 m_\chi \mu_p^2 } 
          B_\SI(\ER) \left[ M_A^2 \right],\\
\kappa^{}_\SD(\ER,m_\chi)
&=& 
\label{structure}
     N_T \frac{\rho_\chi M_A}{2 m_\chi \mu_p^2 } B_\SD(\ER) 
        \left[\frac43 \frac{J+1}{J}
        \left( \langle S_p \rangle \cos\theta
              +\langle S_n \rangle \sin\theta   
       \right)^2\right]  ,\\
B_{\SI,\SD}(\ER) 
&=& \nonumber
        \frac{\langle v \rangle}{\langle v^2 \rangle}
        F^2_{\SI,\SD}(\ER)I(\ER). 
\end{eqnarray}
        Here $\langle {S}_{p(n)} \rangle $ is the spin of the proton 
        (neutron) averaged over all nucleons in the nucleus $A$.
        The dimensionless integral $I(\ER)$ 
        is dark-matter-particle velocity distribution correction, 
        which reduces the rate for large enough momentum transfer:
$$
I(\ER)= \frac{ \langle v^2 \rangle}{ \langle v \rangle }
 \int_{x_{\min}} \frac{f(x)}{v} dx 
    = \frac{\sqrt{\pi}}{2}
\frac{3 + 2 \eta^2}{{\sqrt{\pi}}(1+2\eta^2)\erf(\eta) + 2\eta e^{-\eta^2}}
                [\erf(x_{min}+\eta) - \erf(x_{min}-\eta)],
$$
        where we assume that in our Galaxy rest frame 
        WIMPs have the Maxwell-Boltzmann velocity distribution and
        use the dimensionless Earth speed with respect to the halo 
        $\eta=1$,\  
        $\displaystyle x_{\min}^2 = 
        \frac{3}{4}\frac{M_A\ER}{\mu^2_A{\bar{v}}^2}$,
        $\displaystyle \mu_A = \frac{m_\chi M_A}{m_\chi+ M_A}$.
        The error function is 
        $\displaystyle \erf(x) = \frac{2}{\sqrt{\pi}}\int_0^x dt e^{-t^2}$.
        The velocity variable is the dispersion $\bar{v}\simeq 270\,$km$/$c.
        The mean WIMP velocity 
        ${\langle v \rangle} = \sqrt{\frac{5}{3}} \bar{v}$.
        For the WIMP mass density in our Galaxy 
        the value $\rho_\chi=0.3$~GeV$/$cm$^3$ is used. 
        We also assume both form-factors 
        $F^2_{\SI,\SD}(\ER)$ in the simplest Gaussian form following
\cite{Ellis:1988sh,Ellis:1991ef}.
         In particular, this allows rather simple formulas 
(\ref{structure}) to be used, which are suitable for our comparative 
	 consideration.
 	The total direct detection rate $R(\eth, \emx)$
	can be obtained by integrating differential rate 
(\ref{Definitions.diff.rate}) 
	over the recoil energy interval from the threshold energy $\eth$ 
	to the maximal energy $\emx$.
	To accurately estimate the event rate $R(\eth, \emx)$
	one needs to know a number of quite uncertain 
	astrophysical and nuclear structure parameters
	as well as the very specific characteristics of an experimental setup
	(see, for example, discussions in 
\cite{Bernabei:2003xg,Bernabei:2003za}). 
        In this paper it is enough to assume these uncertainties to be 
	almost the same for all target materials considered. 
	Furthermore,
        one should calculate cross sections 
        $\sigma^{}_\SD$ and $\sigma^{}_\SI$ within the 
	framework of, for example,  
        some SUSY-based theory or take them from experimental data.
	In 
\cite{Bednyakov:2004be}  
        both $\sigma^{}_\SD$ and $\sigma^{}_\SI$ have already been
	calculated within a phenomenological SUSY model  
	whose parameters are defined 
	directly at the electroweak scale 
	(see e.g.
\cite{Mandic:2000jz,Bergstrom:1996cz,Gondolo:2000fh,Bergstrom:2000pn,%
Bednyakov:2003wf,Bednyakov:2002dz,Bednyakov:2002js,%
Bednyakov:2002mb,Bednyakov:2000he,Bednyakov:2002ng,%
Bednyakov:1999vh,Bednyakov:1997jr,Bednyakov:1997ax,Bednyakov:1994qa}). 
        This effective scheme of the MSSM is called the effMSSM in 
\cite{Bottino:2000jx}, and later the 
	low-energy effective supersymmetric theory (LEEST) in 
\cite{Ellis:2003ry,Ellis:2003eg}. 
	The effMSSM parameter space is determined by entries of the mass 
	matrices of neutralinos, charginos, Higgs bosons, sleptons and squarks.
	The list of free parameters includes 
	$\tan\beta$, the ratio
	of neutral Higgs boson vacuum expectation values; 
	$\mu$, the bilinear Higgs parameter of the superpotential;
	$M_{1,2}$, soft gaugino masses; 
	$M_A$, the CP-odd Higgs mass; 
	$m^2_{\widetilde Q}$, $m^2_{\widetilde U}$, $m^2_{\widetilde D}$ 
	($m^2_{\widetilde L}$, $m^2_{\widetilde E}$), 
	squark (slepton) 
	mass parameters squared for the 1st and 2nd generation;        
	$m^2_{\widetilde Q_3}$, $m^2_{\widetilde T}$, $m^2_{\widetilde B}$ 
	($m^2_{\widetilde L_3}$, $m^2_{\widetilde \tau}$), 	
	squark (slepton) mass parameters squared 
	for the 3rd generation; 
	$A_t$, $A_b$, $A_\tau$, soft trilinear 
	couplings for the 3rd generation.
	The third gaugino mass parameter $M_3$ defines the 
	mass of the gluino in the model and is 
	determined by means of the GUT assumption $M_2 = 0.3\, M_3$.
	The intervals of the randomly scanned MSSM parameter space in 
\cite{Bednyakov:2004be}  
	were narrowed to fit the DAMA-inspired  
	domain of the lower masses of the LSP ($m_\chi < 200$~GeV). 
	The current experimental 
	upper limits on sparticle and Higgs masses
	from the Particle Data Group as well as 
        the limits on the rare $b\rightarrow s \gamma$ decay 
	have been imposed. 
	For each point in the MSSM parameter space (MSSM model) 
	the relic density of the light neutralinos 
	$\Omega_{\chi} h^2_0$ 
	was evaluated with the code 
\cite{Bednyakov:2002dz,Bednyakov:2002js,Bednyakov:2002ng} based on 
        DarkSUSY
\cite{Gondolo:2000ee} with the allowance for 
	all coannihilation channels with 
	two-body final states that can occur between neutralinos, charginos,
	sleptons, stops and sbottoms
	as long as their masses are $m_i<2m_\chi$.
	Two cosmologically interesting regions are considered in 
\cite{Bednyakov:2004be}. 
        One is 
	$0.1< \Omega_\chi h^2  < 0.3$ and the other is 
	the WMAP-inspired region $0.094< \Omega_\chi h^2  < 0.129$ 
\cite{Spergel:2003cb,Bennett:2003bz}.
         Further details can be found in 
\cite{Bednyakov:2004be}.

\section{Results and discussion} 
\subsection{Calculations in effMSSM}
	Integrating the differential rate 
(\ref{Definitions.diff.rate}) 
	from the recoil energy threshold $\eth$ 
	to some maximal energy $\emx$
        one obtains the total detection rate $R(\eth, \emx)$
	as a sum of the SD and SI terms:
\begin{equation}
\label{Definitions.total.rate}
        \displaystyle
R(\eth, \emx) = R_{\SI}(\eth, \emx) + R_{\SD}(\eth, \emx) = 
      \int^{\emx}_{\eth} d\ER \kappa^{}_\SI(\ER,m_\chi)\,\sigma_\SI
    + \int^{\emx}_{\eth} d\ER \kappa^{}_\SD(\ER,m_\chi)\,\sigma_\SD.
\end{equation}         
        In 
\cite{Bednyakov:2004be}
	estimations of the ideal total expected rate $R(0,\inf)$ 
	for WIMPs with $M^{}_{\rm WIMP}< 200$~GeV$/c^2$
	in a $^{73}$Ge detector are obtained
	within the effMSSM. 
	Here, with $\sigma^{}_\SD$ and $\sigma^{}_\SI$ already calculated in 
\cite{Bednyakov:2004be}, new estimates of the integrated event rate 
	$R(\eth, \emx)$ for WIMP masses smaller than 200~GeV$/c^2$
	are obtained for a number of DM targets.
	For definiteness, the recoil energy threshold
	$\eth=5$~KeV (and sometimes $\eth=10$~KeV)
	with the maximal energy $\emx=50$~keV are used.
	
\begin{figure}[p] 
\begin{picture}(100,100)
\put(-25,-5){\includegraphics{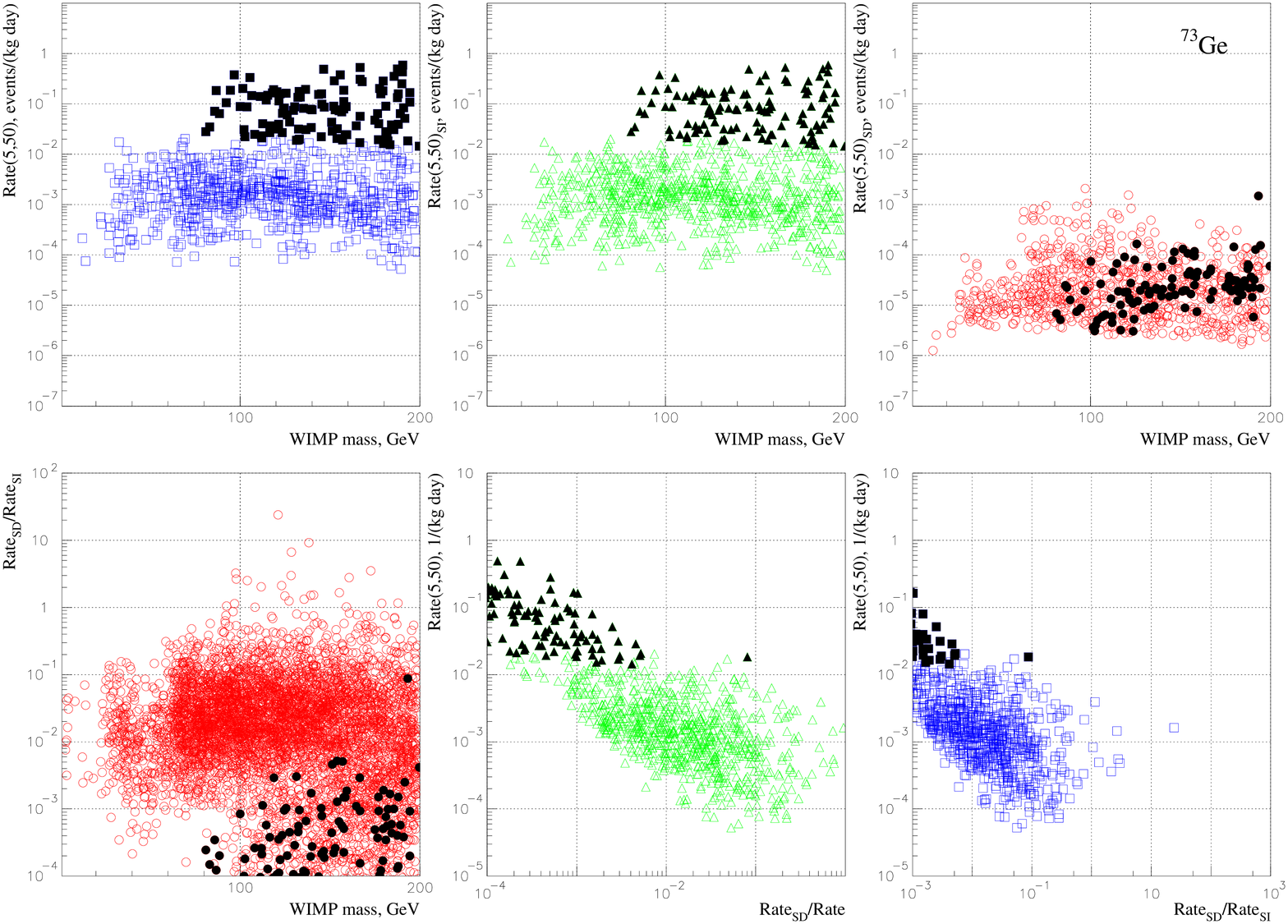}}
\end{picture}
\caption{Expected in $^{73}$Ge total $R(5,50)$, 
         SI and SD event rates $R(5,50)_\SI$ and $R(5,50)_\SD$ (upper panel)
         as well as the ratio $R(5,50)_\SD/R(5,50)_\SI$ (lower left panel)
         as functions of WIMP mass. 
	 Correlations between the total rate $R(5,50)$ 
	 and the SD fraction in $R(5,50)$ is given in the lower middle panel.
	 Correlations between the total rate $R(5,50)$ 
	 and the ratio $R(5,50)_\SD/R(5,50)_\SI$
	 is given in the lower right panel.
	 Open symbols correspond to the 
	 WMAP constraint $0.094< \Omega_\chi h^2  < 0.129$.  
	 Closed symbols give rates with an extra DAMA constraint
	 $1\cdot10^{-7}~{\rm pb}<\sigma^p_\SI<3\cdot 10^{-5}~{\rm pb}$. 
\label{Rates-Ge73}}
\begin{picture}(100,110)
\put(-25,-5){\includegraphics{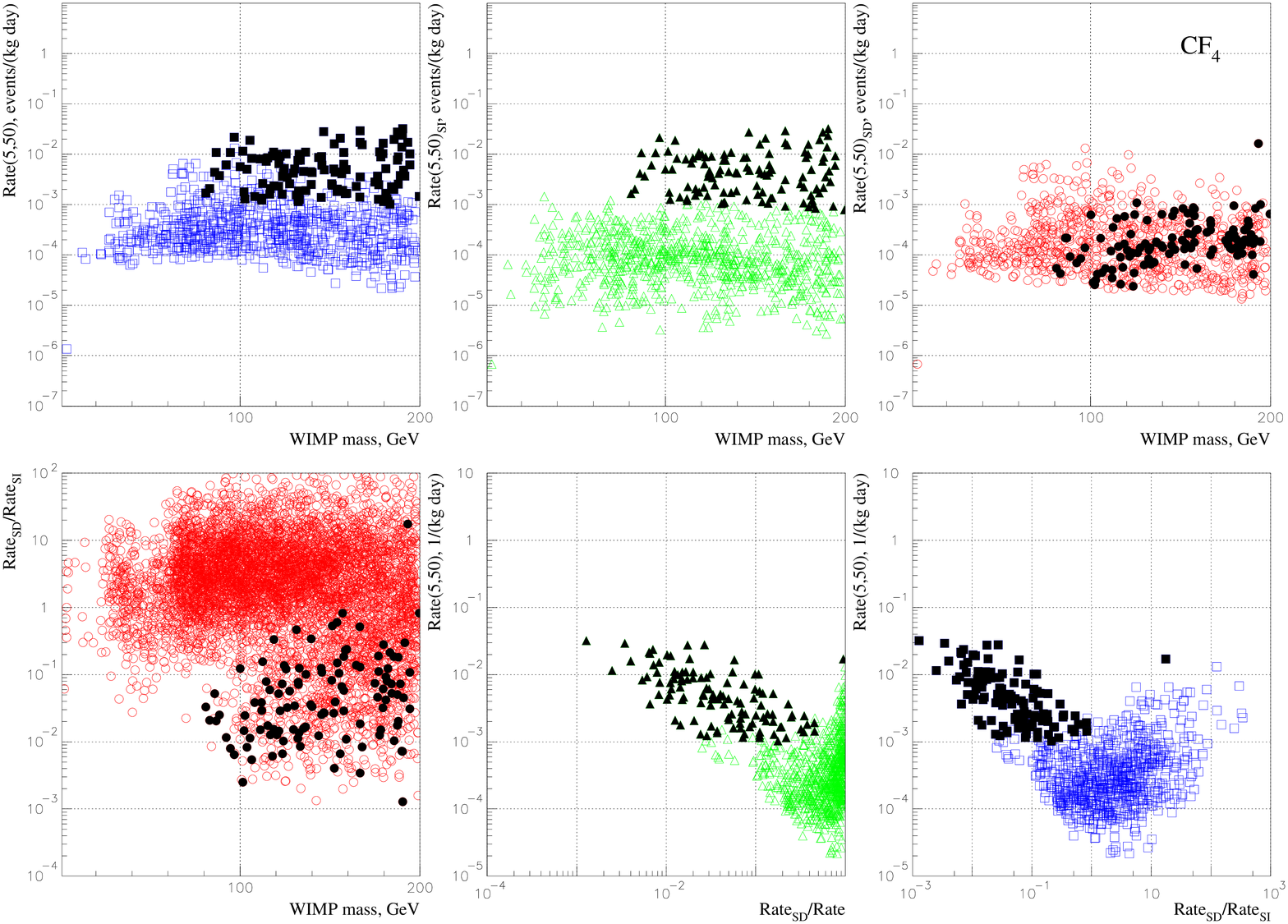}}
\end{picture}
\caption{The same as in Fig.~\ref{Rates-Ge73}, but for CF$_4$.
\label{Rates-CF4}}
\end{figure}

\begin{figure}[!h] 
\begin{picture}(100,105)
\put(-27,-5){\includegraphics{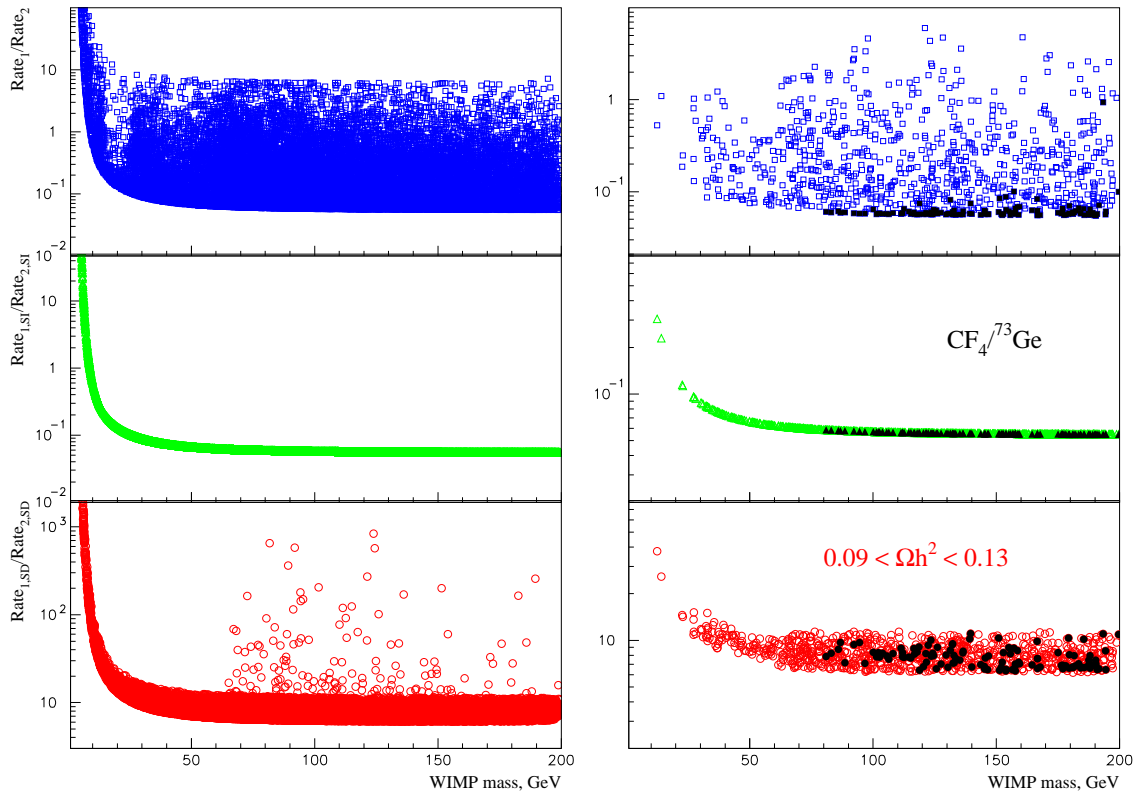}}
\end{picture}
\caption{Ratios $R(5,50)_{{\rm CF}_4}/R(5,50)_{^{73}{\rm Ge}}$
         of the total (top panel), SI (middle) and SD (bottom) rate
	 as functions of WIMP mass.
         No relic density constraint is imposed in the left column.
	 Open symbols in the right column correspond to the 
	 WMAP constraint $0.094< \Omega_\chi h^2  < 0.129$.  
	 Filled symbols give these ratios for the rates obtained 
	 with an extra DAMA SI constraint,
	 $1\cdot10^{-7}~{\rm pb}<\sigma^p_\SI<3\cdot 10^{-5}~{\rm pb}$. 
\label{Ratios2Ge73-CF4}
}
\end{figure}

\begin{figure}[p] 
\begin{picture}(100,100)
\put(-25,-5){\includegraphics{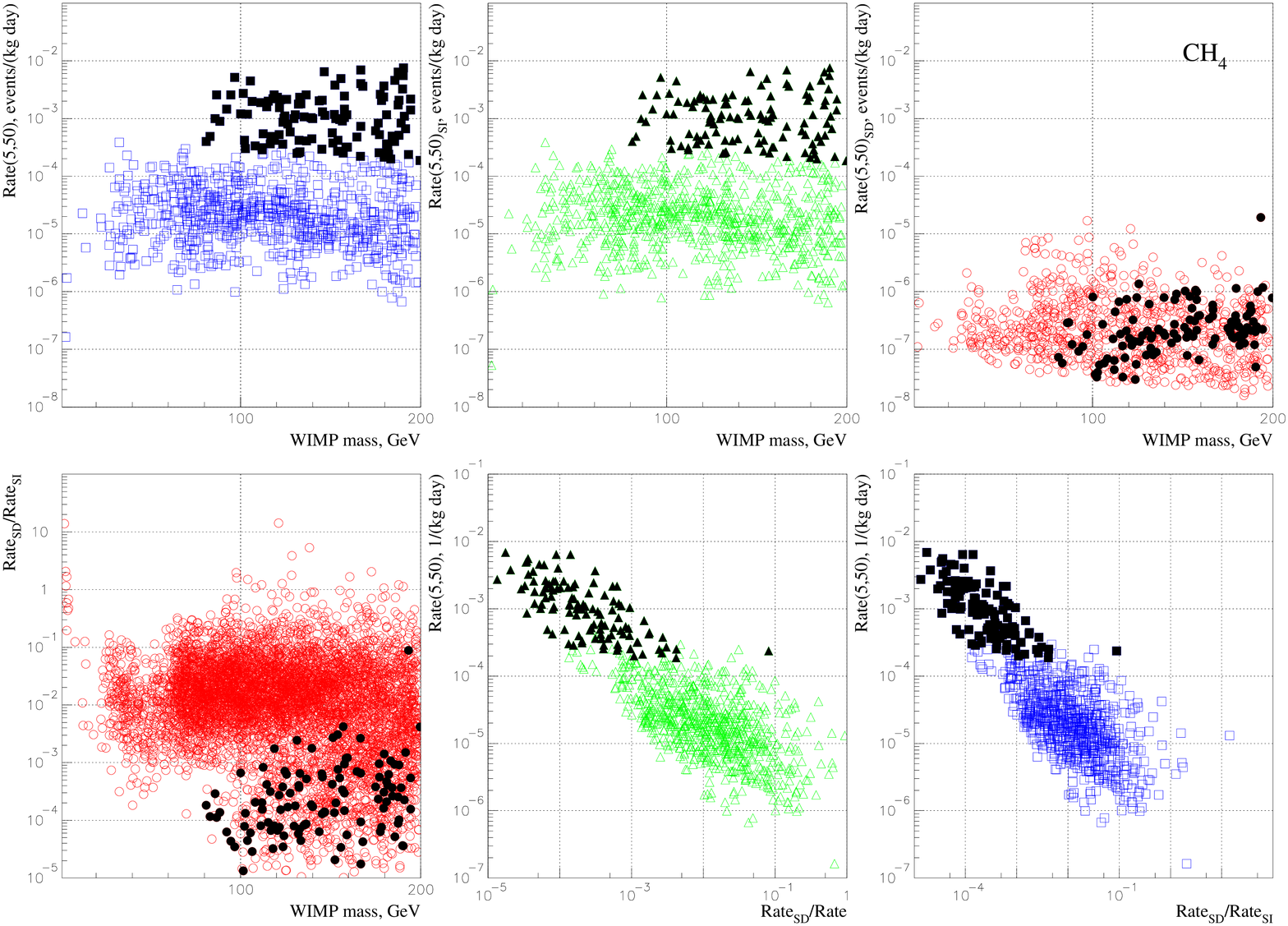}}
\end{picture}
\caption{
  Expected in the CH$_4$ target the total $R(5,50)$, SI and SD event rates 
              $R(5,50)_\SI$ and $R(5,50)_\SD$ (upper panel)
              as well as the
              ratio $R(5,50)_\SD/R(5,50)_\SI$ (lower left panel)
              as functions of WIMP mass. 
	      Correlations between the total rate $R(5,50)$ 
	      and the SD fraction in $R(5,50)$ is given in the lower middle panel.
	      Correlations between the total rate $R(5,50)$ 
	      and the ratio $R(5,50)_\SD/R(5,50)_\SI$
	      is given in the lower right panel.
	      Open symbols correspond to the 
	      WMAP constraint $0.094< \Omega_\chi h^2  < 0.129$.  
	      Closed symbols give rates with an extra DAMA SI constraint,
	      $1\cdot10^{-7}~{\rm pb}<\sigma^p_\SI<3\cdot 10^{-5}~{\rm pb}$. 
\label{Rates-CH4}
} 
\begin{picture}(100,110)
\put(-25,-5){\includegraphics{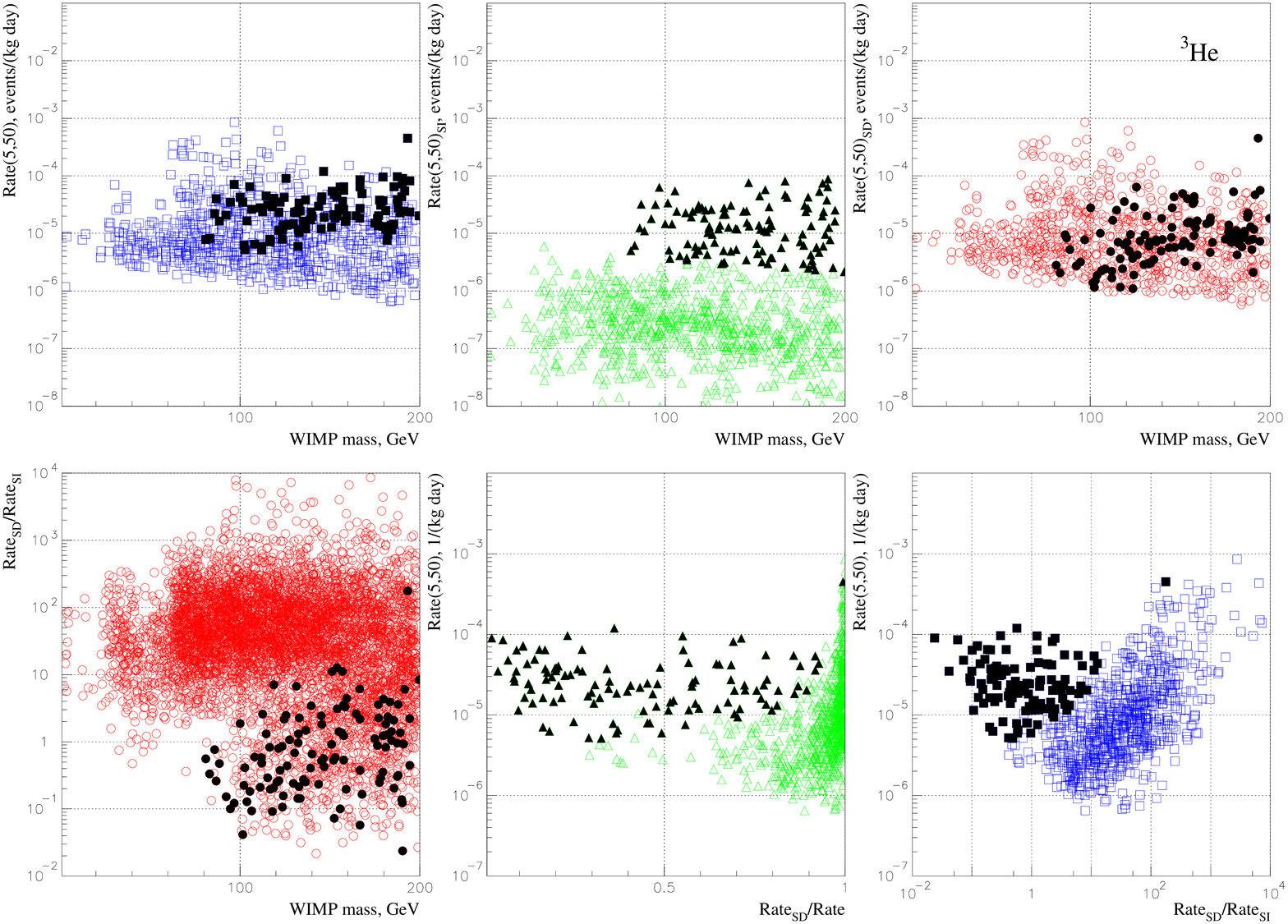}}
\end{picture}
\caption{The same as in 
Fig.~\ref{Rates-CH4}, but for $^3$He.
\label{Rates-He3}
}
\end{figure}

\begin{figure}[!h] 
\begin{picture}(100,105)
\put(-27,-5){\includegraphics{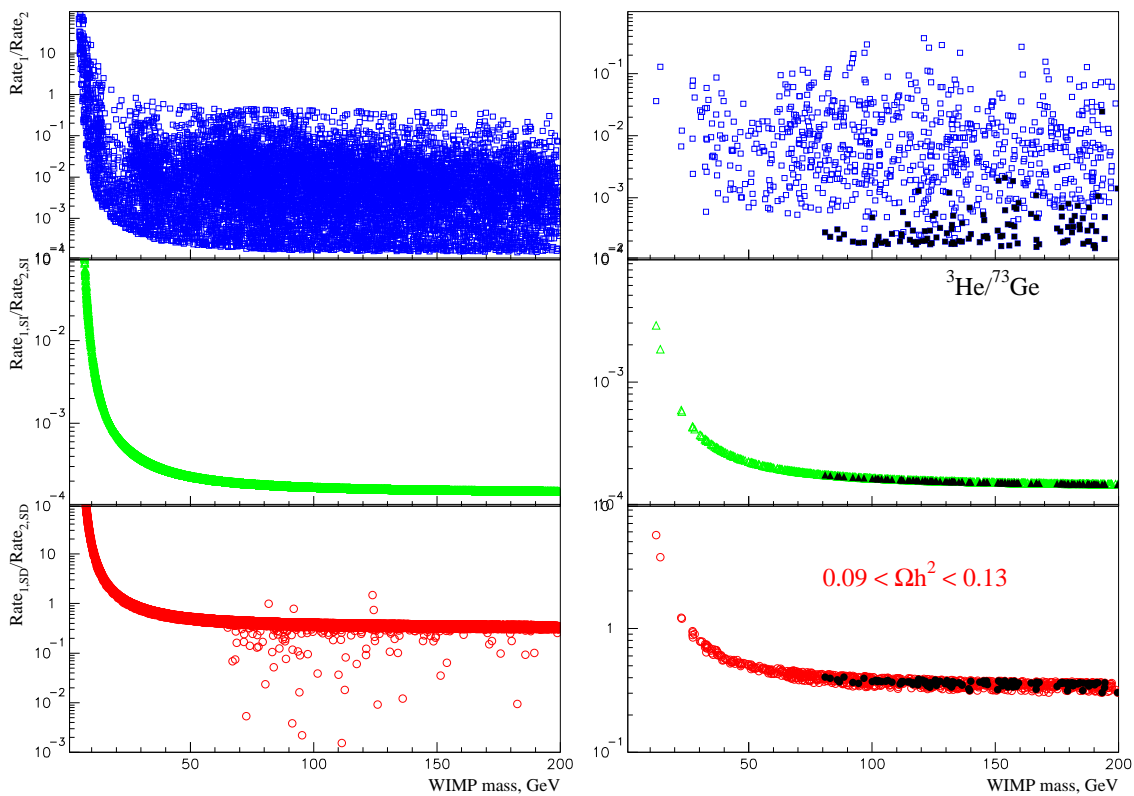}}
\end{picture}
\caption{Ratios $R(5,50)_{^3{\rm He}}/R(5,50)_{^{73}{\rm Ge}}$
         of the total (top panel), SI (middle) and SD (bottom) rate
	 as functions of WIMP mass.
         No relic density constraint is imposed in the left column.
	 Open symbols in the right column correspond to the 
	 WMAP constraint $0.094< \Omega_\chi h^2  < 0.129$.  
	 Filled symbols give these ratios for rates obtained 
	 with an extra DAMA SI constraint,
	 $1\cdot10^{-7}~{\rm pb}<\sigma^p_\SI<3\cdot 10^{-5}~{\rm pb}$. 
\label{Ratios2Ge73-He3}
}
\end{figure}

\begin{figure}[!p] 
\begin{picture}(100,105)
\put(-27,-5){\includegraphics{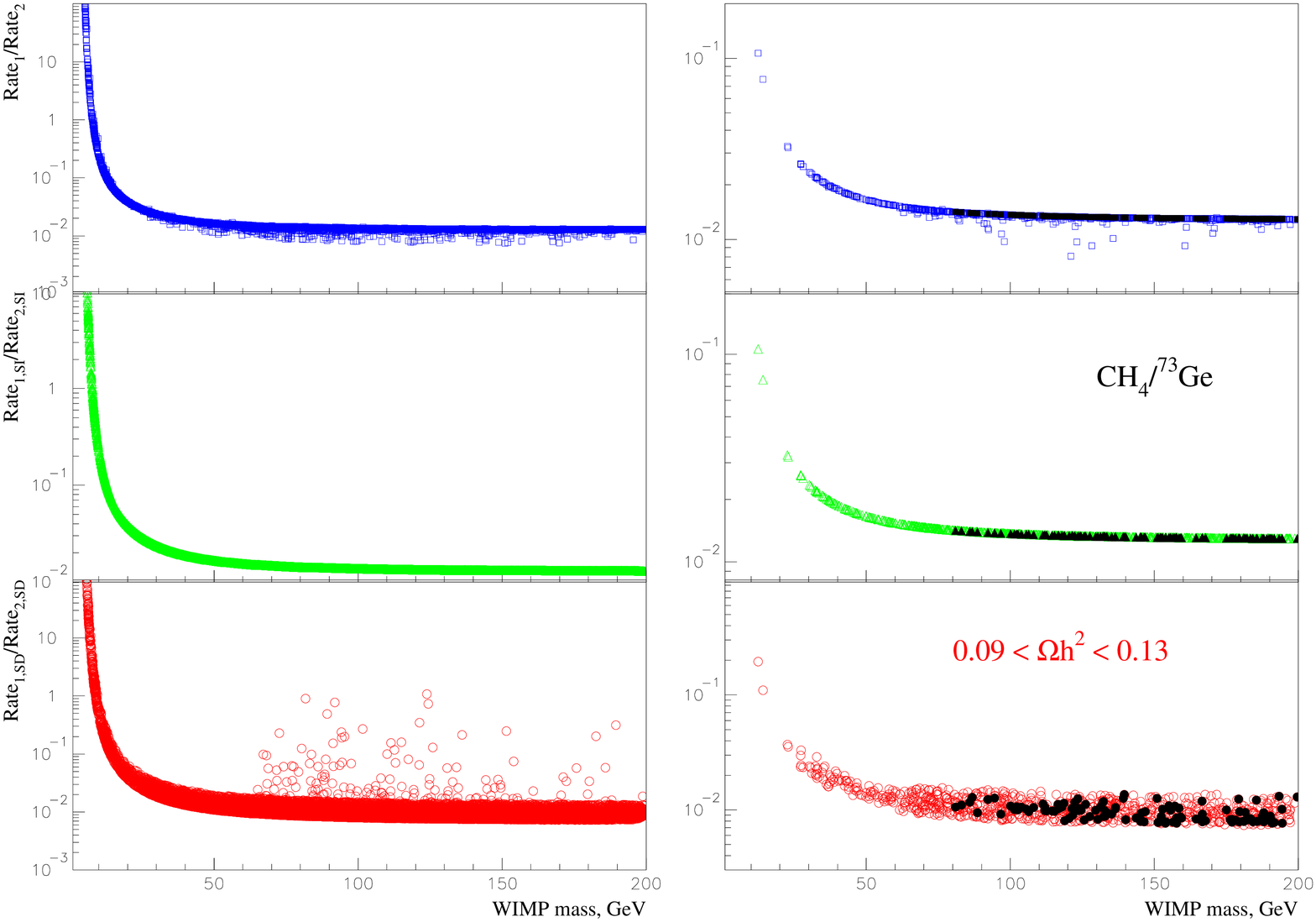}}
\end{picture}
\caption{Ratios $R(5,50)_{{\rm CH}_4}/R(5,50)_{^{73}{\rm Ge}}$
         of the total (top panel), SI (middle) and SD (bottom) rate
	 as functions of WIMP mass.
         No relic density constraint is imposed in the left column.
	 Open symbols in the right column correspond to the 
	 WMAP constraint $0.094< \Omega_\chi h^2  < 0.129$.  
	 Filled symbols give these ratios for rates obtained 
	 with an extra DAMA SI constraint,
	 $1\cdot10^{-7}~{\rm pb}<\sigma^p_\SI<3\cdot 10^{-5}~{\rm pb}$. 
\label{Ratios2Ge73-CH4}
}
\begin{picture}(100,105)
\put(-27,-5){\includegraphics{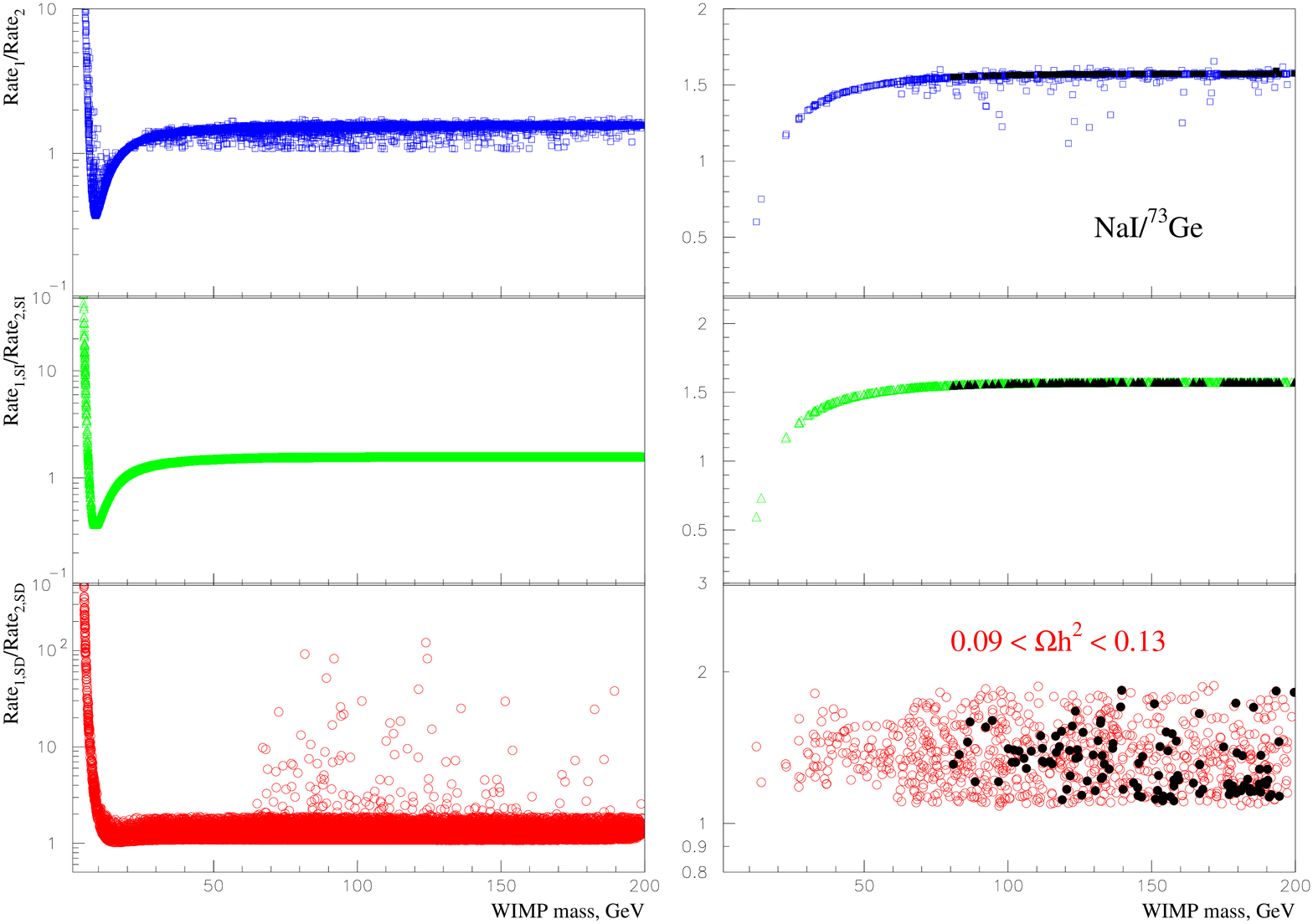}}
\end{picture}
\caption{The same as in Fig.~\ref{Ratios2Ge73-CF4}, but  for NaI and $^{73}$Ge.
\label{Ratios2Ge73-NaI}
}
\end{figure}

	The calculated event rates $R(5,50)$ and the rate ratios for
	different targets 
	are depicted as scatter plots in 
Figs.~\ref{Rates-Ge73}--\ref{Ratios2Ge73-NaI}. 
        For example, in 
Fig.~\ref{Rates-Ge73} one can see total, $R(5,50)$, 
	spin-independent, $R(5,50)_\SI$ and spin-dependent $R(5,50)_\SD$ 
	event rates expected in $^{73}$Ge-target together with 
        their ratio $R(5,50)_\SD/R(5,50)_\SI$
	as functions of the WIMP mass. 
	It is always interesting to trace some interplay between 
	SD and SI contributions to the total event rate.
	To this end correlations between the total rate $R(5,50)$ 
	and its SD fraction as well as 
        correlations between the total rate $R(5,50)$ 
	and the SD-to-SI ratio $R(5,50)_\SD/R(5,50)_\SI$ are also given
in Fig.~\ref{Rates-Ge73}.  
        All open symbols in the figures correspond to the case when 
	the WMAP constraint on the relic neutralino density 
	$0.094< \Omega_\chi h^2  < 0.129$ is taken into account.
	The relevant filled symbols show the 
	rates which one would expect if an extra DAMA constraint, 
	$1\cdot10^{-7}~{\rm pb}<\sigma^p_\SI<3\cdot 10^{-5}~{\rm pb}$,
	is imposed on SI cross sections. 
	It is seen that in the last case (filled symbols)
	the SI rates are at least two orders of magnitude larger than
	the SD one and the large 
	total rate values ($R>0.01$~events/day/kg)
	are saturated only by the SI interactions. 
	If one ignores these filled symbols (i.e. the DAMA-inspired
	extra constraint
	$1\cdot10^{-7}~{\rm pb}<\sigma^p_\SI<3\cdot 10^{-5}~{\rm pb}$),
	then the SD contribution does not look very suppressed 
	and the SD contribution alone can saturate the total event rate, but 
	only 
	when the rate itself is rather small ($R \approx 0.001$~events/day/kg).
	These features take place for all heavy enough targets,
	therefore the corresponding figures 
	for NaI, CsI, and Xe target are not given.
	
	It is well known that a fluorine-containing target is the best one
	for detection and measurement of the spin-dependent WIMP-nucleus 
	interaction (see e.g.
\cite{Divari:2000dc,Jungman:1996df}).
Figure~\ref{Rates-CF4} shows that 
        the SD rate in CF$_4$ is indeed the biggest one and 
	for a large number of points (MSSM models)
	the SD contribution dominates.
        Nevertheless it is also seen that it is not correct to completely 
	ignore the SI contribution to the total expected rate in the fluorine
	target
\cite{Bednyakov:2005qp,Bednyakov:2004be} because the 
        SI rate is almost the same as the SD one.
        Furthermore, at a current level of the DM detector sensitivity,
	when the DAMA-inspired large SI contributions 
	are not yet completely excluded (filled symbols
in Fig.~\ref{Rates-CF4}),
        the SD contribution in CF$_4$, $R(5,50)_{\SD}$, 
	is smaller than the SI one, $R(5,50)_{\SI}$.
	The ratios of the total, SI and SD rates in the CF$_4$ and 
	$^{73}$Ge targets are presented in 
Fig.~\ref{Ratios2Ge73-CF4} as a function of WIMP mass.

	 The ratios in the left panel of 
Fig.~\ref{Ratios2Ge73-CF4} correspond to the rates calculated 
         without any constraint on the relic density of 
	 neutralinos in the effMSSM.
	 The increase of these ratios at very low WIMP masses 
	 reflects better sensitivity of fluorine to smaller
	 WIMP masses than that of germanium. 
	 Open symbols in the right panel of 
Fig.~\ref{Ratios2Ge73-CF4} depict these ratios when the 
	 WMAP constraint $0.094< \Omega_\chi h^2  < 0.129$ is imposed 
	 on the calculated neutralino relic density. 
	 Filled symbols gives these ratios for the rates obtained 
	 with an extra DAMA SI constraint,
	 $1\cdot10^{-7}~{\rm pb}<\sigma^p_\SI<3\cdot 10^{-5}~{\rm pb}$. 
	 The sensitivity of  CF$_4$ to the SD WIMP interaction 
	 is about ten times as large as the SD sensitivity of $^{73}$Ge. 
	 At the same time, the CF$_4$ sensitivity 
	 to the SI WIMP interaction is
	 less than 0.1--0.05 of the SI sensitivity of $^{73}$Ge. 
	 As a result, the total expected 
	 rate in $^{73}$Ge is a bit larger than in a very spin-sensitive
	 CF$_4$ target.
	 The right panel in 
Fig.~\ref{Ratios2Ge73-CF4} shows that
	 the relic density WMAP and extra DAMA constraints 
	 make this conclusion stricter.
	 The expected total rate in a heavy enough $^{73}$Ge target 
	 is about ten times as large as 
	 the total expected rate in the CF$_4$ target.

\enlargethispage{\baselineskip}

Figures~\ref{Rates-CH4} and \ref{Rates-He3}
     show the total, SD and SI expected event rates in the CH$_4$ and $^3$He 
     targets which contain the lightest nonzero-spin nuclei, interesting for 
     direct DM search.
     Absolute values of the total, SD, and SI rates in these materials are 
     visibly smaller than in the fluorine-containing CF$_4$ and 
     germanium targets
(Figs.~\ref{Ratios2Ge73-He3}--\ref{Ratios2Ge73-CH4}),
     especially when the extra WMAP (left panel) and DAMA (filled symbols) 
     constraints are imposed.

     For all detectors with heavy enough target  mass 
     (NaI, CsI, Xe, etc) the absolute values of the total, SI and SD rates 
     look very similar with the only possible exception in the domain 
     of very low-mass WIMPs (when a target contains some light isotope 
     like, for example, Na in the NaI target).
     In 
Figure~\ref{Ratios2Ge73-NaI} a set of NaI-to-$^{73}$Ge rate ratios, 
     $R(5,50)_{\rm NaI}/R(5,50)_{^{73}{\rm Ge}}$,
     is given for illustration of the behavior.
     All depicted ratios are of the order of unity. 
     Only for very low WIMP masses (less than 10 GeV/$c^2$)
     and with the WMAP constraint neglected 
     the rates in NaI start to clearly dominate 
     over the rates in $^{73}$Ge due to 
     kinematically preferable WIMP interaction with Na.  
     This low-WIMP-mass growth of the ratios is 
     absent in the CsI and Xe targets.
     Furthermore, for these  materials the rates in 
     $^{73}$Ge dominate in the low-mass WIMP region.

\subsection{Calculations with DAMA constraints}
         In the previous section  the WIMP-nucleon $\sigma^{}_{\SD}$ 
         and $\sigma^{}_{\SI}$ cross sections which enter event rate 
(\ref{Definitions.total.rate}) 
	 were taken from theoretical calculations in the effMSSM  
\cite{Bednyakov:2004be}.
\begin{figure}[!h] 
\begin{picture}(100,95)
\put(-5,-13){\includegraphics{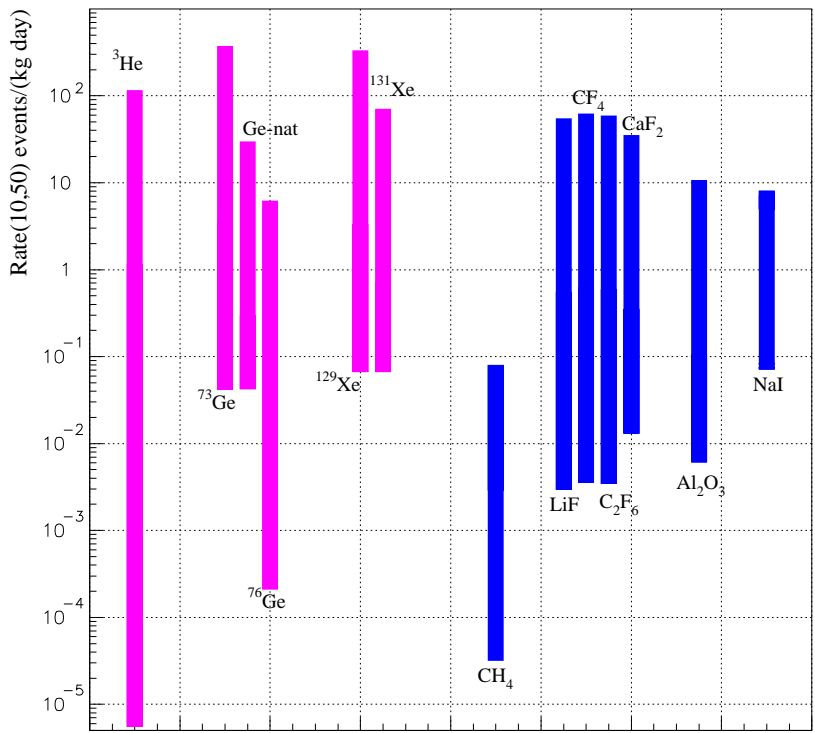}}
\end{picture}
\caption{Variations of expected event rates, $R(5,50)$, for 
a number of targets followed from the DAMA-allowed cross sections 
  $\sigma^{}_{\SD}$ and $\sigma^{}_{\SI}$.
	 Targets with nonzero-spin nuclei from the  
	 odd-neutron (odd-proton) group model are given
	 in the left (right) part of the figure. 
\label{DAMA-R}
}
\end{figure} 
         Another source of these cross sections is, for example,  
	 the DAMA experiment
\cite{Bernabei:2003za}.
\begin{figure}[!h] 
\begin{picture}(100,98)
\put(-5,-12){\includegraphics{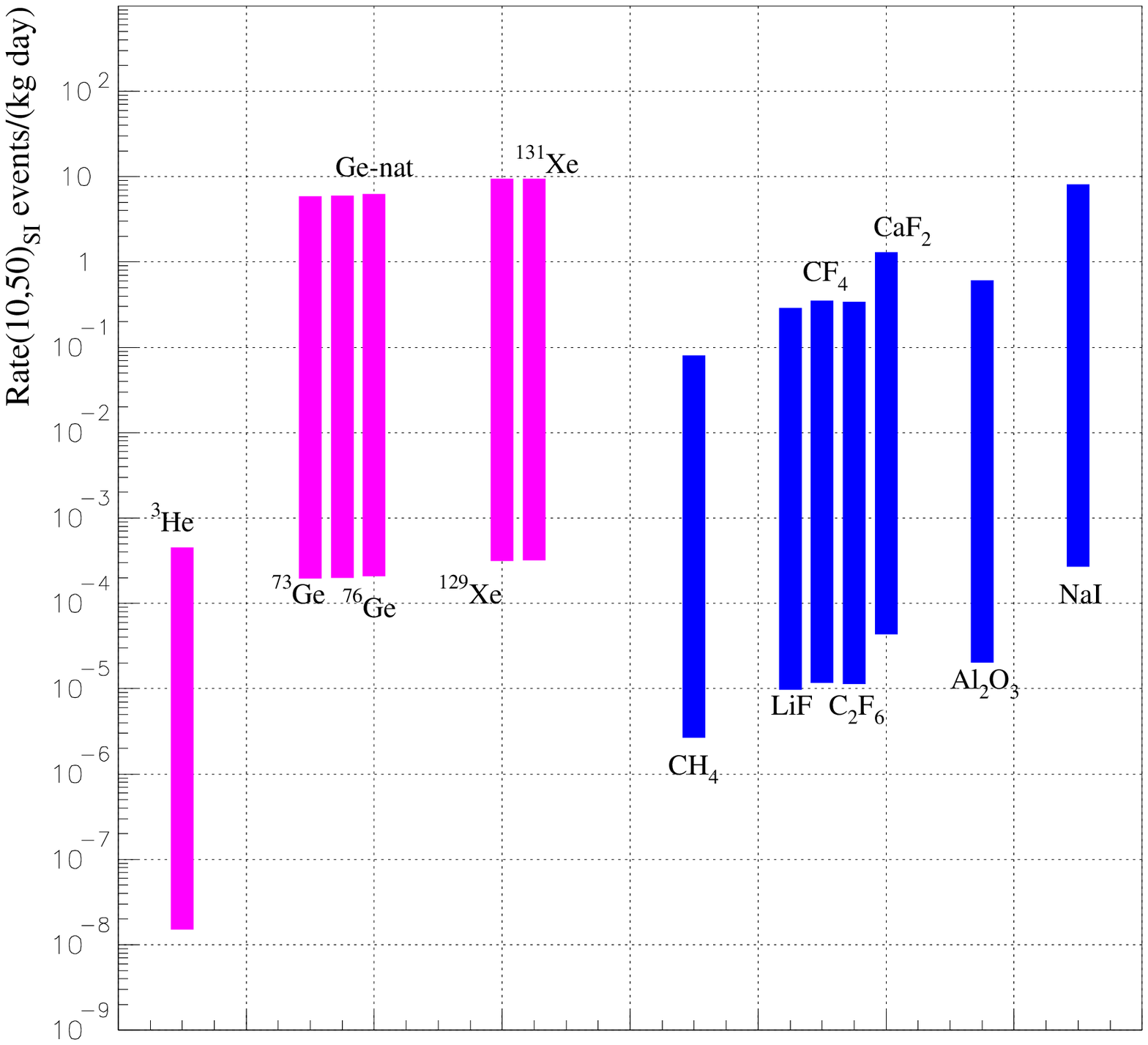}}
\end{picture}
\caption{Variations of expected spin-independent contributions   
  to the event rate, $R(5,50)_{\SI}$, in a number of targets
  followed from the DAMA-allowed cross sections 
  $\sigma^{}_{\SD}$ and $\sigma^{}_{\SI}$.
\label{DAMA-RSI}
}
\begin{picture}(100,100)
\put(-5,-12){\includegraphics{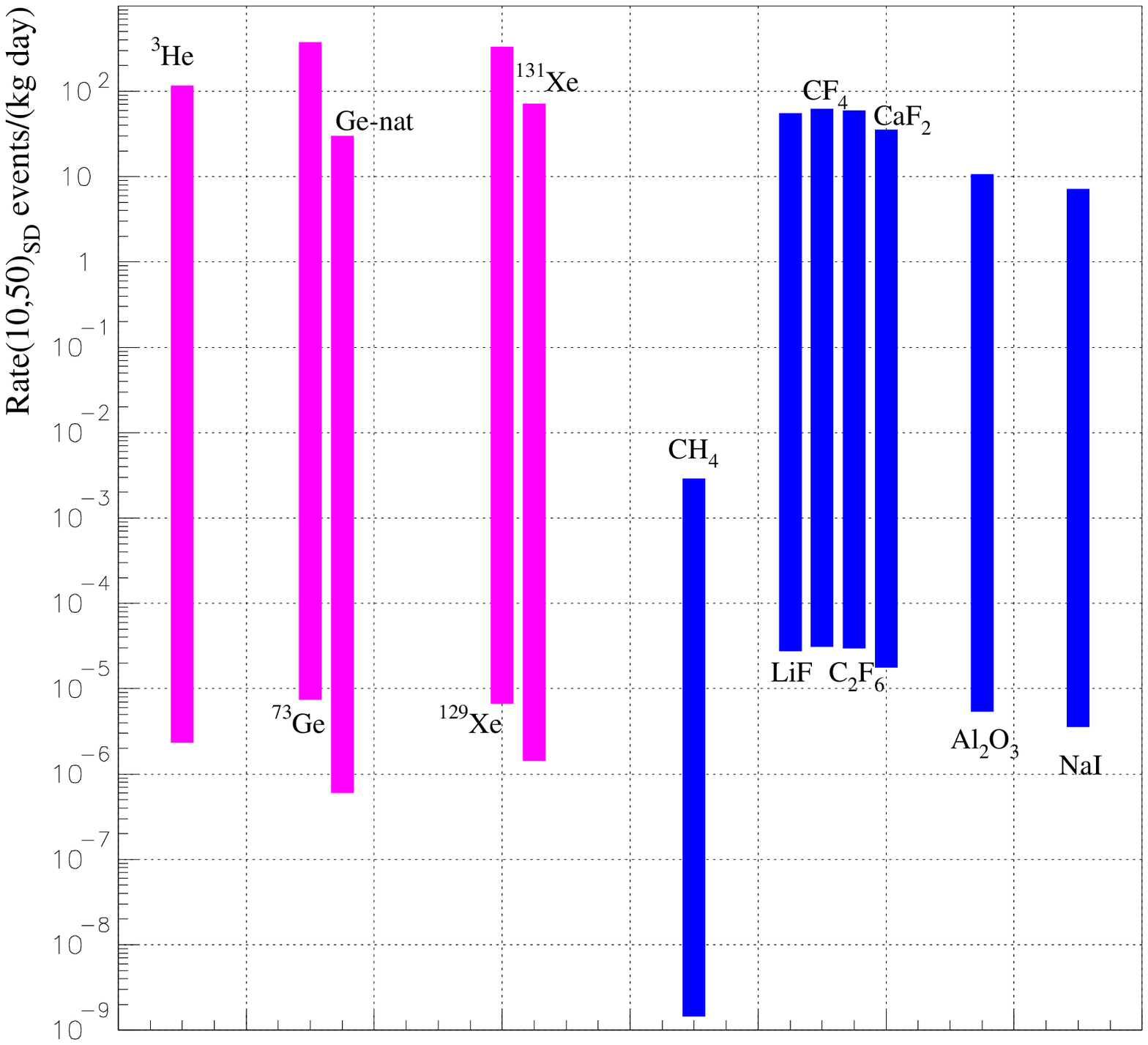}}
\end{picture}
\caption{The same as in 
Fig.~\ref{DAMA-RSI}, but for the 
spin-dependent contributions  $R(5,50)_{\SD}$.
\label{DAMA-RSD}
}
\end{figure} 
         With formulas
(\ref{Definitions.diff.rate})--(\ref{Definitions.total.rate}) 
         the DAMA(NaI) constraints
         on $\sigma^{}_{\SD}$ and $\sigma^{}_{\SI}$ from 
\cite{Bernabei:2003za} 
         can be phenomenologically transformed into the 
         allowed regions for the detection rates with other targets.
	 The results of these recalculations are given in 
Figs.~\ref{DAMA-R}--\ref{DAMA-RSD} 
         for the total, SI and SD expected rates $R(10,50)$
	 in a number of representative materials for a DM detector. 
	 Here the threshold of 10~keV is used. 
	 The values of the expected rate can vary within 
	 the columns without any conflict with 
	 the DAMA-allowed 
	 $\sigma^{}_{\SD}$ and $\sigma^{}_{\SI}$ regions.
	 The left (right) parts of these figures contain 
	 rate restrictions for 
	 the odd-neutron (odd-proton) group model nonzero-spin nuclei
(see e.g.
\cite{Bednyakov:2004xq}).
          Some of the highest rate values, for example in $^{73}$Ge,
	  are already excluded by measurements
\cite{Klapdor-Kleingrothaus:2005rn}.
          In particular, from 
Figs.~\ref{DAMA-R}--\ref{DAMA-RSD} 
          one can see that all fluorine-containing 
	  targets (LiF, CF$_4$, C$_2$F$_6$ and CaF$_2$, etc)
	  have almost the same sensitivity to the both SD and SI
	  WIMP-nucleus interactions.
	  Among all materials considered a detector with 
	  a $^{73}$Ge, $^{129}$Xe, or NaI target
	  has better prospects to confirm or to reject 
	  the DAMA result 
	  due to the largest values
	  of the lower bounds for the total rate
	  ($R(10,50)>0.06-0.08$~events/day/kg). 
	  If, for example, one ignores the SI WIMP interaction
(Fig.~\ref{DAMA-RSI}),
	  then all materials have almost the same 
	  prospects to detect DM particles
	  with the only exception of CH$_4$.

\section{Conclusion} 
	Expected event rates for a number of dark matter target materials
	are calculated in the effective low-energy minimal
	supersymmetric standard model (effMSSM), provided the
	lightest neutralino is the dark matter 
	Weakly Interacting Massive Particle (WIMP).
	The results obtained are based on previous
	evaluations of the neutralino-proton (neutron) 
	spin and scalar cross sections 
	for the neutralino masses $m_\chi < 200$~GeV/$c^2$
\cite{Bednyakov:2004be}.
	The performed calculations allow direct comparison of sensitivities 
	of different dark matter setups to the
	WIMPs expected from the measurements of the 
	DAMA experiment.
	In particular, it is shown that detectors with 
	a $^{73}$Ge, $^{129}$Xe, and NaI target
	have better prospects to confirm or to reject the DAMA result. 

	It is worth noting, that to get very 
	accurate predictions for the event rate 
        one has to take into account a number of quite uncertain 
        astrophysical and nuclear parameters and 
        specific features of a real setup.
        We considered only a simple spherically symmetric isothermal 
        WIMP velocity distribution 
\cite{Drukier:1986tm,Freese:1988wu}
        and do not go into detail of any possible
        and in principle important 
        uncertainties (and/or modulation effects) 
        of the Galactic halo WIMP distribution 
\cite{Kinkhabwala:1998zj,Donato:1998pc,Evans:2000gr,%
Green:2000jg,Copi:2000tv,Ullio:2000bf,Vergados:2000cp}.
        For simplicity we use the Gaussian 
        scalar and spin nuclear form-factors from 
\cite{Ellis:1993ka,Ellis:1991ef}.
         We believe it is relevant 
	 for our comparative study because the very influence    
         of the factors is suppressed in the rate ratios.

\smallskip
        The authors thank the RFBR (02--02--04009) 
	and the JINR--Slovak Republic program for support.

\clearpage

\providecommand{\href}[2]{#2}\begingroup\raggedright\endgroup

\end{document}